# An Adaptable Fast Matrix Multiplication (AFMM) Algorithm: Going Beyond the Myth of "Decimal-War"

Niraj Kumar Singh[1], Soubhik Chakraborty[2*] and Dheeresh Kumar Mallick[1]

[1]Department of Computer Science & Engineering, B.I.T. Mesra, Ranchi-835215, India
[2]Department of Applied Mathematics, B.I.T. Mesra, Ranchi-835215, India
*email address of the corresponding author: soubhikc@yahoo.co.in (S. Chakraborty)

**Abstract:**
In this paper we present an adaptable fast matrix multiplication (AFMM) algorithm , for two nxn dense matrices which computes the product matrix with average complexity $T_{avg}(n) = \mu' d_1 d_2 n^3$ with the acknowledgement that the average count is obtained for addition as the basic operation rather than multiplication which is probably the unquestionable choice for basic operation in existing matrix multiplication algorithms. Here $d_1$ and $d_2$ are the densities (fraction of non-zero elements) of the pre and the post factor matrices only and $\mu'$ is the expected value of the non zero elements of the post factor matrix. Remembering the fact that a single addition operation is much cheaper (however, this factor may differ from one machine to another) than a single multiplication operation, our algorithm finds the product matrix without using a single multiplication operation. The replacement of multiplications by additions has several significant and interesting aspects as it adds a non-determinism even to a problem which otherwise is considered to be deterministic! It can be argued that for inputs trivial as well as non trivial, AFMM algorithm can beat Strassen's algorithm for matrix multiplication.

## 1. Introduction:

Not everything that can be counted counts and not everything that counts can be counted.
Albert Einstein (1879-1955)

The classical version along with its major variants, where fixing the instance size (n) fixes all the computing operations is deterministic as it produces the same output when it is re-run for the identical inputs. In this paper we present an adaptable matrix multiplication algorithm, for two nxn dense matrices. Our algorithm computes the resultant matrix in average time $O(\mu' d_1 d_2 n^3)$ when analyzed mathematically. This discussion is further followed by its parameterized analysis as well. Here $d_1$ and $d_2$ are the densities (fraction of non-zero elements) of the pre and the post factor matrices only.

Since the time of Strassen's discovery [1] in 1969, several other algorithms for multiplying two nxn matrices of real numbers in $O(n^\alpha)$ time with progressively smaller constants α have been invented. The fastest algorithm so far is that of Virginia Williams [2], with its efficiency in $O(n^{2.3727})$. The decreasing values of the exponents have been obtained at the expense of increasing complexity of these algorithms. Because of large multiplicative constants, none of them is of practical use [3].

The deterministic nature of response remained unchanged even when we switched to more sophisticated algorithms for this problem. The only positive aspect in the whole development is interesting only from theoretical point of view. But when it comes to practice, especially for average case analysis, it is the actual machine time which matters more than the theoretical predictions based on the assumption of pivotal ,i.e., dominant operation(s) present in the code.

Irrespective of their sophistication, a major commonality among the present approaches is their parameter independence nature. That is, in terms of number of basic operations performed, they all are almost unaffected of the value of inputs. This article is a major breakthrough in this direction due to its adaptive nature to the input values present in the matrices. Remembering that a single addition operation is much cheaper (however, it may differ from one machine to another) than a single multiplication operation we have introduced an algorithm which solves this problem without using even a single multiplication operation. The replacement of multiplications by additions has several significant and interesting aspects as it adds non-determinism even to a problem which otherwise is considered to be deterministic [4]. With suitable modifications even the classical matrix multiplication algorithm qualifies to be a potential candidate for parameterized complexity analysis.

## 2. Adaptable Matrix Multiplication (AFMM) Algorithm

Below we present an adaptable matrix multiplication (AFMM) algorithm for the multiplication of two dense square matrices. Let X and Y be pre and post factor square matrices of size nxn with densities $d_1$ and $d_2$ respectively. Here Z is the resultant (product) matrix initialized with all zeroes. In case-A, the pre factor matrix X consists of real number values generated randomly using some probability distribution. Whereas, matrix Y consists of integer values whose expected value is assumed as µ. In case-B, the values inside the pre and post factor matrices swap their properties.

As the method outlined below deals with multiplication of two dense matrices we recommend adjacency matrix as data structure for storing matrix data.

```
Case-A:
     REAL X, Z, Base
     INTEGER Y, I, J, K, Rep-factor, p
1 ....WHILE I ← 1 to N DO
2 ..........WHILE K ← 1 to N DO
3 ................. Base ← X [I, K]
4 ................ IF (Base EQUALS 0)
5 ...................... Redirect control to the beginning of 'K' loop
6 ................. WHILE J ← 1 to N DO
7 ...................... Rep-factor ← Y [K, J]
8 ...................... Z [I, J] ← Z [I, J] + $\sum_{p=1}^{\mu''} Base$ // µ'' = floor (µ)

Case-B:
```

```
    REAL Y, Z, Base
    INTEGER X, I, J, K, Rep-factor, p

1 ....WHILE I ← 1 to N DO
2 ..........WHILE K ← 1 to N DO
3 ................. Rep-factor ← X [I, K]
4 ................. WHILE J ← 1 to N DO
5 ....................... Base ← Y [K, J]
6 .......................IF (Base EQUALS 0)
7 ....................... ....... Redirect control to the beginning of 'K' loop
8 ....................... Z [I, J] ← Z [I, J] + $\sum_{p=1}^{\mu''}$ Base  // $\mu''$ = floor ($\mu$)
```

## 2.1 Proof of Correctness

We present the following arguments in favor of the correctness of AFMM algorithm.

Point (1): AFMM computes correct result (Case-A is discussed).

The computation is done in row-major fashion. A selected X [I, K] (base) element is processed Y [K, J] number of times, and this activity is repeated n times for J=1 to n. This individual activity is basically summation of the base element Y [K, J] number of times. Once J values exhaust, the inner most loop (J loop) is terminated and the whole process is iterated for incremented value of K. Following these lines, upon exhausting the K values, we have final result for $I^{th}$ row of the resultant matrix Z [I, J].

Whole of the above process is repeated for next I values (2 to n) as well and finally when all the $I'^s$ get exhausted we get the final result in the form of matrix Z [I, J].

Also, as the "rep-factor" value is assumed to be an integer, in statement-8 the upper value to summation is rightly justified.

Point (2): AFMM stops eventually.

As the depth of each of the three loops is bounded from above by the dimension of pre and post factor matrices it always eventually terminates.

## 2.2 Theoretical Analysis (Counting the expected number of additions)

Since addition is the only key operation identified in the algorithm, we obtain the theoretical complexity in terms of expected number of additions. The expected number of additions E(A), is a function of both the dimension of the two matrices as well as the expected value of elements (denoted as E(X) = $\mu$) inside the post (pre in case of case-b) factor matrix Y (X).

Let $\mu'$ be the expected value of non-zero elements in the post factor matrix Y (Case A) and which is known to the experimenter in advance and $\mu$ be the mean over the entire post factor matrix elements. A little calculation shows that µ equals to µ′$d_2$. The expected number of addition $E(A) = n\{n^2d_1\}µ$. Here the factor $n^2d_1$ corresponds to each of the non-zero elements of the pre factor matrix and the factor n corresponds to the frequency of each of such elements contributing to the final result. Substituting µ′$d_2$ for µ we get $E(A) = µ′d_1d_2n^3$, which is the expected number of additions. Trivially, when product of the two density factors is kept fixed at $n^{-1}$ we obtain a quadratic run time, provided the factor µ' is reasonable!

## 2.3 Empirical Analysis through Statistical Bound Estimate (Empirical-O)

This section includes empirical results. The observed mean time (in second(s)) was noted in table (1). Average case analysis was done by directly working on program run time which can be used in estimating the weight based statistical bound over a finite range by running computer experiments [5], [6]. This estimate is called empirical-O [4], [7]. Here time of an operation is taken as its weight. *Weighing permits collective consideration of all operations into a conceptual bound which we call a statistical bound in order to distinguish it from the count based mathematical bounds that are operation specific*. The credibility of empirical-O depends on the design and analysis of our special computer experiment in which time was response. See [4] for more insight.

*System specification:* All the computer experiments were carried out using PENTIUM 1600 MHz processor and 512 MB RAM.

### 2.3.1 Parameterized Complexity Analysis

This section does a systematic measurement of our proposed algorithm over differing parameter values. Through our experimentation we have re-verified the parameter independence nature of standard 'ijk' and 'ikj' algorithms [8]. However, the replacement of multiplications by additions has induced non determinism to AFMM algorithm. And hence in this respect it demands for a comprehensive parameterized analysis for this algorithm.

The observed mean time over sufficient number of readings for each specified 'n' value is recorded as is given in table (1). The parameters $d_1$ and $d_2$ are 1/3 and ½ respectively for µ′ = 1 through 7, and 1/5 and 2/5 respectively for µ′ equal to 14 and 21. These data essentially correspond to case (A) of the proposed algorithm.

Table (1): observed mean time in second(s)

| N↓ | ijk | ikj | AFMM µ'=1 | AFMM µ'=3 | AFMM µ'=5 | AFMM µ'=7 | AFMM µ'=14 | AFMM µ'=21 |
|---|---|---|---|---|---|---|---|---|
| 250 | 0.411667 | 0.328 | 0.1752 | 0.2784 | 0.3316 | 0.3498 | 0.3188 | 0.2874 |
| 500 | 2.718667 | 2.32 | 0.918 | 1.5406 | 1.943 | 2.331 | 1.9122 | 1.9874 |
| 750 | 10.235 | 7.75533 | 2.8475 | 4.9718 | 6.38 | 7.6826 | 6.3218 | 6.4328 |
| 1000 | 25.125 | 18.43233 | 6.67175 | 11.8064 | 15.073 | 18.2756 | 14.9158 | 15.3374 |
| 1250 | 49.437 | 35.75 | 13.0897 | 23.0705 | 29.489 | 35.594 | 29.172 | 29.9267 |
| 1500 | 85.531 | 61.688 | 22.5625 | 40.078 | 50.9295 | 61.718 | 50.3515 | 52.2 |
| 1750 | 135.906 | 98.515 | 36.0235 | 63.2655 | 80.6235 | 97.6955 | 80.281 | 81.9335 |
| 2000 | 205.422 | 147.078 | 53.375 | 94.6875 | 120.8515 | 145.8985 | 119.516 | 122.491 |

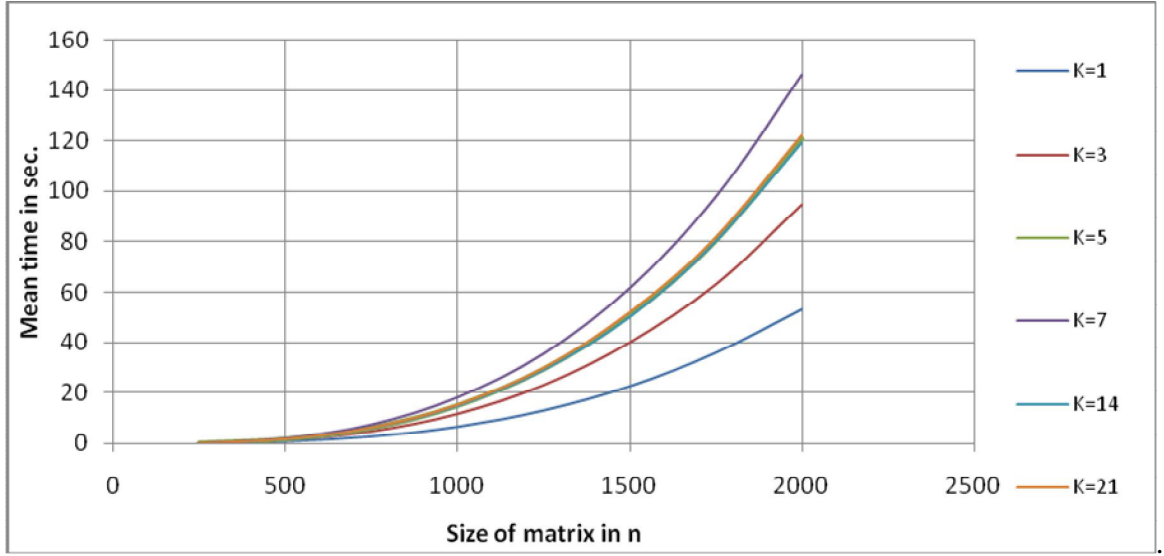
Fig (1) Plot for AFMM algorithm on various specified parameters

The plot for data in table (1) is given in fig (1). Here treat constant k as µ' which is the mean of non-zero elements in matrix Y. Assuming the run time measurement for 'ikj' algorithm as the baseline, it is observed that for the specified parameter values the run time of AFMM algorithm is reduced by around 64 percent when the expected value is 1. It is reduced by 19 and 17 percent when the expected value is 14 and 21 respectively.

## 3 Conclusions

Amongst the algorithms for multiplying two matrices, Strassen's algorithm has an edge over others for sufficiently large (practically feasible) matrices [3]. Still with suitable constraints (while maintaining both the matrices dense), as mentioned in the main article, we can always generate inputs for which AFMM beats Strassen's. Trivially by keeping the product $d_1d_2$ at 1/n we can hope for a quadratic performance from AFMM which is never expected from that of Strassen's irrespective of the type of input. Even for non-trivial inputs with sufficiently low values of $d_1d_2$ product we expect better performance from AFMM.

Fortunately the assumptions made in this paper regarding the values in either of pre/post factor matrices can be generalized for arbitrary valued dense matrices. Our work towards this generalization is in progress and we are hopeful of getting some exciting results out of it.